# Cosmological Natural Selection Revisited
## Some Remarks on the Conceptual Conundrum and Possible Alleys

**Rainer E. Zimmermann**
IAG Philosophische Grundlagenprobleme,
Fachbereich 1 der Universität,
Nora-Platiel-Str.1, D – 34127 Kassel /
Clare Hall, UK – Cambridge CB3 9AL [1] /
Lehrgebiet Philosophie,
Fachbereich 13 AW, Fachhochschule,
Lothstr. 34, D – 80335 Muenchen[2]
e-mail: pd00108@mail.lrz-muenchen.de

**Abstract**

Following the selection metaphor as introduced by Lee Smolin in his 1997 book with respect to a possible model of the reproduction of Universes, this model is being re-constructed utilizing the strict analogical form of the metaphor chosen. It is asked then for the genotypal level associated with the primarily phenotypal model, and it is asked in particular where the information processing mechanism of that cosmological sort of selection could actually be found. It is argued that massive black holes in the centres of galaxies may play this important role. Some consequences on black holes in general are discussed then pointing to the necessity to actually revisit the concepts of virtual and actual black holes also.

## 1. Introduction

Following the book publication by Lee Smolin [1], I have discussed his model of cosmological natural selection twice elsewhere ([2], [3]), pointing to a number of inconsistencies of the model and possible variants which may be helpful in solving one or the other problem with it. After these preliminary discussions we are now capable of re-formulating Smolin's conception in a generalized way by means of applying directly the somewhat corrected version of the analogy between physics and biology actually being put forward. The advantage is to concretely establish *selection* as an intrinsically universal principle indeed. The Universe in the *facon de parler* of Smolin's is then nothing but the individual sample whose type is the analogue of the *phenotype*. Hence, we shift from the concept of selection to what we call *superselection* in order to point to the fact that the principle invoked here acts upon *networks* (populations) of Universes which have to be classified according to their type.

---
[1] Permanent addresses.
[2] Present address.

The latter is being defined in the sense of the excess productivity of black holes actually achieved, and is a measure therefore for the possible number of „offspring" of Universes which may be eventually produced. We have thus a population of Universes which can be stratified into sub-populations reflecting the various phenotypes. (The question as to whether a population may contain Universes which are not space-times, simply due to epistemological reasons, we leave aside for the time being. [4]) These phenotypes are involved in competition among each other. Superselection acts thus onto these phenotypes according to the environmental structure in which the sub-populations participate. Hence, what we need is the equivalent of an environment, some *intercosmic mediator*.

But let us first have a look onto the interior „ecology" of a *single* Universe: A generic Universe consists of a hierarchic structure of galaxies (supercluster, cluster, local groups, individual samples of galaxies). It is indeed the *stellar ecology* of galaxies which explicitly determines the production rate of black holes. Hence, also a galaxy can be visualized as an analogue to local ecosystems containing populations of phenotypes. In this case the types are *stellar* types plus a mediator which is given in terms of the *interstellar matter*. The Universe appears thus as a hierarchically organized population of populations of local ecosystems. But if the stellar type (described by the spectral class of stars) is a local version of a phenotype, what is then the associated genotype?

Remember that biology is essentially organized in terms of three structural and functional levels which we can call the molecular, cellular, and organismic levels, respectively. The last one is directly associated with the phenotype. The molecular level coincides with the level for which this expression has been introduced in the first place, namely the chemical level. So what is the cellular level? This would be exactly that level on which the „cosmogenetic" coding is being stored, in a place which would be the equivalent of a *cell nucleus*.

A short intermediate remark to the utilization of Smolin's selection metaphor: Indeed, at the occasion of a short discussion of the topic some time ago [5], Smolin has not shown much enthusiasm when being confronted with the arguments unfolded above. Instead, he seemed to be pointing to the fact that this metaphor chosen by him would only be applicable in a somewhat limited way and could not directly be transferred to a physical terminology. But here we have to object that the metaphor is apperently of *some* use after all. However, this being the case, it has to be developed with all its consequences. And the latter are to be examined carefully with a view to their consistency. If they turn out ot be inconsistent, the principle has to be given up. But there cannot be a mere *semi-metaphor* in sufficiently general terms. Hence, a complete discussion of the principle introduced is what is being asked for.

## 2. A First Ansatz

Hence, what we do is to start in a straightforward manner from the original analogy which is itself referring to the organismic structure of organization as known from biology: Therefore, phenotypes are types of organisms which are themselves structured in a molecular way and constituted on that molecular level by means of the respective genotype. It is important here to remember oncemore that we will not expect that we actually deal with explicitly biological quantities after all, but that instead we will have to deal with physical quantities only. Insofar the metaphor stays nothing but a metaphor. On the other hand, we know of course, that biological systems are again nothing but chemical and thus complex physical systems themselves. Hence, we can legitimately argue to found the further analogy onto the relative orders of magnitudes being actually involved: The individual body of human beings e.g. represents a sample of the human phenotype and thus constitutes the organism on the macroscopic level in terms of biological aspects. Let us say that the human body contains some $10^2$ organs, which are composed of different sorts of webs (according to the various fields of sensory perceptions, food processing, motion, reproduction, and so forth). With roughly $10^{14}$ cells per organism this comes to about $10^{12}$ cells per organ. (The molecular mass of human cells is at about $10^{15}$.) Hence, the ratio of the „highest" to the „lowest" level of organization is about $10^{10}$ or $10^{12}$, according to whether one refers to the organs or the whole organism.

On the other hand, the Universe contains about $10^{11}$ galaxies of which each for itself contains $10^{11}$ stars on the average. The galaxies are organized in clusters and superclusters with about $10^6$ and $10^9$ galaxies, respectively. Hence, it is intuitively clear, within the chosen analogy, to actually identify the level of superclusters with the level of organs, and the level of galaxies with the level of cells. The „scalelessly" mediated orders of magnitude among clusters refer then to the levels of different biological webs. Note that in biology, organs are usually not only classified according to their *structure*, but also according to their *function*. That is, organs have a specific function within the whole organism. Moreover, there must be a *fine tuning* of functions among each other, because any malfunction has to be compensated for a while for not endangering the organism altogether. In the long run, any malfunction will put the organism at a decisive risk. Nothing else is the case with the physical functions of the constituents of the Universe.

If now galaxies are the equivalent of cells, what would be formally equivalent to the nucleus of a cell? The point is that the biological analogy can be carried quite far in the case of galaxies, because like cells they indeed perform specific functions of some „metabolism" and can therefore be visualized as regulatory systems. (This is in fact what we also claim when speaking about body cells of an organism.) The sort of metabolism mentioned here is however one of a cosmic exchange of matter, where the important things happen in places we cannot observe very well: in the dark spaces in between spiral arms of galaxies. Because this is where stars are being „cooked" which „cook" themselves other matter then and so forth. It is from then on that they become visible to us. Hence, the complete galaxy gains the connotation of a self-organizing (massively parallel) computational system, very much like the biological cell.

The cells utilize the information which is contained in the genetic structure and is actually stored in the cell's nucleus, and produce proteins from it. The DNA is very much like a hard disk, while the proteins are a sort of RAM. [6] We can claim something similar with respect to the galaxy, provided we think of its „nucleus" also as a comparable system for storing and processing information. Indeed, there is a literal nucleus to most galaxies which is represented by a *central massive black hole*. In other words: It is such a black hole which would be the adequate place to „store the laws of nature in some symbolical manner" as Smolin is himself asking for. (He actually disputes the possibility of such a storing of symbolical information though. But that is a point where doubts should be in order.) Obviously, the next question is *how* the relevant information could be stored in black holes. Unfortunately, the interior of black holes is not well-known until now. [7] However, massive black holes in the centres of galaxies as they have gained renewed relevance since the beginning of 2000 when the Chandra satellite observatory did its first impressive series of X-ray photos, could well be visualized as that „gates towards a new physics" as Martin Rees has introduced them recently. [8] The characteristic „careers" of black holes have been studied in detail some time ago by Kip Thorne among others. [9] Many aspects recovered here correspond nicely to what Smolin proposes in his approach.

These aspects by the way imply also the absence of the fundamental categories of space and time which we have discussed elsewhere. [10] Thorne e.g. compares them within the context of a black hole with a piece of wood soaked with water coming into a fire. Time (the water) evaporates and the remaining space (dry wood) becomes ashes (Wheeler's quantum foam). [11] And indeed, within this physics of the quantum foam we should look for the analogy of (cosmic) nucleotides. Not in the foam itself (because the storing of information should not be subject to random fluctuations), but as its first product. The latter might turn out as *spin networks* in fact. [12] Hence, the clarifying of the analogy is closely related to the problem of developing a TOE.

The same is actually true for what we can visualize as „intercellulary" space: The respective region in between the galaxies carrying the reserves for the latter's metabolistic functions is typically visualized as an almost perfect void. But even if this should be true, there is still gravitation. Also within galaxies, the patterns of interstellar matter recently observed could be understood as resulting from exterior „environmental" inputs establishing explicit forms of harmony. [13]

## 3. Computational Aspects

But there is still another point: The internal structure of the Universe alone cannot be decisive for superselection, because we deal with populations of Universes in the first place. That we can retrace the basic structure of selection also within the individual sample and all of its substructures is only an epiphenomenon of the underlying model's scale independence (which is actually a necessary condition for our evolutionary principle in order to adapt it to self-organized criticality in the sense of the Santa Fe school). Hence, we have to differ between *internal* selection in the sense of Smolin (within a given Universe subject to its stellar ecology) on the one hand, and *external* selection of types of Universes (superselection) on the other hand. Obviously, the one has to be fitted to the other. In fact, what we can do is actually to visualize internal selection as a projection (more precisely: a projected image) of superselection. In other words: The biological selection proper shows up then as a mere differentiation of superselection projected onto the planetary ecology. This would indeed shed some light on the question of possible earth-like planets.

But if so, the remaining question is for the „cosmic mediator" already mentioned. It would define the background of Universes on which populations unfold. Hence, it could be visualized as that environment whose structure gives the initial drive for the competition of phenotypes in the first place, because it is only a *finite* environment, one with *restricted* ressources, in other words: a *physically deficient* environment, on which selection can be sufficiently founded. It may be worthwhile to look for this mediating background (which is not a geometrical background as we know it, because we are talking about a region which is beyond Universes, and thus beyond space and time) in the loops of loop quantum gravity itself. In a first instance, the concept of spin networks is relevant here. The problem is that we have to permanently think „in terms of space and time" so that we cannot abstract from these fundamental categories, even if talking about their factual absence. As I have discussed elsewhere [14], spin networks can be visualized as the *boundary* of space-time, and as such they are also the (epistemological) boundary of substance. [15] But the point is that there is no real transition from the one (the world) to the other (substance) and viceversa across that boundary, because substance *is always everywhere* (thus non-locally) underlying the world which we can observe, but which is nothing but a back-projection of substance (which we cannot observe). The question is whether seen under this ontological perspective, it is useful to think of physics as the foundation of biology all the time, while not thinking of the viceversa: that also biology would have something to contribute to the foundation of physics in turn. This may be so because after all, physical theories are being produced by biological living beings according to what the latter can actually perceive. This perception however, together with the thinking applied to it, is primarily biologically constituted. Recently, Louis Kauffman has oncemore reminded on this point and tried to describe a relationship between biology and logic which might be of a specific systematical meaning for what we have said here. [16] Again, there is another parallel to the conception of visualizing the Universe altogether as an emergent computational system which in the case of biology differs from the computers as we know them only in the fact that it is software and hardware at the same time.

## 4. Black Holes

But the essential problem with black holes is as to their eventually becoming „quantum": Originally, the programme was quite straighforward. If a star with a certain minimal mass underwent gravitational collapse, then sooner or later a horizon would form defining the black hole's boundary. And the work concentrated on describing horizon properties. The actual vicinity of the singularity was then cut out from both the space-time manifold and from the discussion, respectively. That was what we learned as students when reading one of our three bibles at the time. [17] Things became more complicated however when Hawking, Bekenstein and others started to discuss quantum aspects of black holes. From the beginning on, the suspicion emerged that classical concepts would be carried over to quantum situations without being really justified. Not that this would have been something new when dealing with quantum physics (indeed it actually established a great deal of distinction for those who worked in „classical" field theories when visualizing the whole Schroedinger programme as some kind of elaborate guessing according to concepts of classical physics). But in the case of black holes it seemed somehow to go too far. All of this despite some very attractive aspects of the new conceptions such as Bekenstein's idea of treating black holes as an analogue to atoms in pre-quantum physics and so forth. However, with the advent of black hole thermodynamics and in particular with the treating of the horizon as a material membrane in the sense of Thorne, things became very interesting, and one was somewhat distracted from the quantum problems lurking underneath. As it turns out now, the very terminology utilized when talking about quantum black holes is covering most of the problems which still have remained. Take the example of the celebrated *Planck mass* which serves as a criterion of talking about a domain where all relevant forces actually meet [18]: It is the frequent changing of the systems of units involved what is diverting from the fact that it is not the Planck mass which is relevant at the Planck level, but instead it is the dimensions involved which are relevant. In other words: If you have a piece of matter with roughly $10^{-8}$ kg of mass, then you do not have normally any problem with quantum fields or quantum gravity as to that – because it is only when you compress this mass to the small Planck length dimensions that it would become theoretically relevant. So it is the *volume* rather than the mass which is important after all. Obviously, when dealing with stellar black holes, you are not dealing with any Planck mass whatsoever, by the very definition above. In fact, we are talking of several solar masses. And the mass remains conserved. What is changing while the star performs a gravitational collapse, is the mass *density* which is enormously increasing because the available volume is decreasing. (And this tells you something about the characteristic gravitation involved which points towards tidal forces rather than to anything else – which also tells something about the entropy change being involved and explains why a black hole is *not* really a time-inverted white hole.)

We know in the meantime from the seminal paper of Rovelli and Smolin and from other papers dealing with its consequences [19] that the volume (as well as the area) of space has to be visualized as a discrete and thus quantized entity. In fact, in case of the volume the quantization is not quite as forward as in the case of the area, but we notice that the spectrum runs proportional to $l_P^3$ as should be expected. This insight has resulted in a number of consequences in the field of quantum information lately. [20] Hence, with respect to what we would like to have in the case of black holes when thinking of what we discussed above, the explicit coupling of gravitation, thermodynamics, and quantum information is very promising after all. Alas, there remains the unsatisfactory situation that the transition of black hole from the classical into the quantum domain is far from clear. The quantization of area and volume of space mentioned above gives some operational rules as to dealing with the computation of quantities at the Planck level. But the important problem here is the following: While undergoing collapse *each* black hole will *necessarily* come into the quantum domain by means of its permanently shrinking volume. And then the question will be how collapse is actually being *halted* at the Planck length, because the true zero-point of length cannot be reached due to the convention of declaring the Planck length to the minimum length which could be possibly attained. Hence, we would conclude that there are no objects with a length which is smaller than the Planck length. But note that a black hole arriving at the Planck length by means of gravitational collapse will not have any Planck mass then! In fact, the mass will be much larger than that which means that the Compton length is actually *undefined*, because it would be smaller than the Planck length for such cases. In other words: while the black hole is entering quantum dimensions, it does not behave like a quantum particle at all. If we cannot ascribe a Compton length to it, then it is not subject to Heisenberg's uncertainty relationship!

In fact, it is quite clear why this should be the case: The reason for this is that within the region of the Compton length, the length scale goes as 1/m, while for the macroscopic region (where the Schwazschild length is of relevance) it goes as m. This is indeed due to the onset of Heisenberg's uncertainty relation. But the question remains why collapsing black holes do not seem to be subject to this transition. On the other hand, postulating virtual black holes which are being created by means of spontaneous fluctuations out of the vacuum does not really help here, because then the question remains whether it is legitimate at all to call such objects black holes if they have not undergone any gravitational collapse.

As it turns out, there are basically two desiderata: One of explicitly demonstrating the transition from a classical volume to a quantum volume displaying the relevant effects such a transition would have for an object undergoing gravitational collapse, and another one of explicitly studying the relationship of length scales at the Planck scale proper as well as defining the state of an object whose Compton length is smaller than its Schwarzschild length, in order to be able to decide whether this could really be called a black hole or not. A first step towards approaching this problem is a recent paper of Paola Zizzi [21]. Further work with more technical explications is in progress. [22] Hence, it is likely to find the information processing mechanism discussed above within the interior of black holes. In the end it is this mechanism on which the concept of cosmological selection is actually founded. But it does not suffice to elaborate further on quantum mechanical details of black holes without being able to say something about the transition from classical to quantum states during gravitational collapse and without clarifying the nature of black-hole like objects at the Planck scale in more detail.

## Acknowledgements

For helpful discussions while in Cambridge I thank John Baez, Julian Barbour, Chris Isham, Lou Kauffman, Mary Hesse, and Lee Smolin. For more recent discussions I thank Paola Zizzi.